\documentclass[twocolumn,preprintnumbers,amsmath,amssymb,superscriptaddress]{revtex4-1}

\usepackage{graphicx}
\usepackage{bm}
\usepackage{color}

\begin{document}


\title{Nonlinear dispersion relation in integrable turbulence}

\author{Alexey Tikan}
\affiliation{Univ. Lille, CNRS, UMR 8523 - PhLAM -
 Physique des Lasers Atomes et Mol\'ecules, F-59000 Lille, France} 
 \affiliation{Present address: Institute of Physics, Swiss Federal Institute of Technology Lausanne (EPFL), CH-1015 Lausanne, Switzerland}
\author{F\'elicien Bonnefoy}
\affiliation{\'Ecole Centrale de Nantes, LHEEA, UMR 6598 CNRS, F-44 321 Nantes, France}
\author{Guillaume Ducrozet}
\affiliation{\'Ecole Centrale de Nantes, LHEEA, UMR 6598 CNRS, F-44 321 Nantes, France}
\author{Gaurav Prabhudesai}
\affiliation{LPS, ENS, CNRS, Univ. Pierre et Marie Curie, Univ. Paris Diderot, F-75 005 Paris, France}
\author{Guillaume Michel}
\affiliation{Institut Jean Le Rond d’Alembert, Sorbonne Université, CNRS, UMR 7190, F-75 005 Paris, France}
\author{Annette Cazaubiel}
\affiliation{Université de Paris, MSC Laboratory, UMR 7057 CNRS, F-75 013 Paris, France}
\author{\'Eric Falcon}
\affiliation{Université de Paris, MSC Laboratory, UMR 7057 CNRS, F-75 013 Paris, France}
\author{Francois Copie}
 \affiliation{Univ. Lille, CNRS, UMR 8523 - PhLAM -
 Physique des Lasers Atomes et Mol\'ecules, F-59000 Lille, France}
\author{St\'ephane Randoux}
 \affiliation{Univ. Lille, CNRS, UMR 8523 - PhLAM -
 Physique des Lasers Atomes et Mol\'ecules, F-59000 Lille, France}
\author{Pierre Suret}
\email{Pierre.Suret@univ-lille.fr}
\affiliation{Univ. Lille, CNRS, UMR 8523 - PhLAM -
  Physique des Lasers Atomes et Mol\'ecules, F-59000 Lille, France}

\date{\today}



\maketitle

\textbf{The concept of \textit{Nonlinear dispersion relation} (NDR) is  used in various fields of Physics (nonlinear optics~\cite{Hutchings1992}, hydrodynamics~\cite{Herbert2010,Taklo2015, Aubourg:16}, hydroelasticity~\cite{deike13}, mechanics~\cite{Cobelli:DispersionRelation:09}, quantum optics~\cite{Carretero2008}, plasma physics~\cite{Benisti2008},...)
to characterize  fundamental phenomena induced by nonlinearity such as wave frequency shift or  turbulence. Nonlinear random waves described by the one-dimensional nonlinear Schr\"odinger equation (1DNLSE) exhibit a remarkable form of turbulence called "integrable turbulence" where solitons play a key role~\cite{Zakharov:09, Walczak:15, Agafontsev:15, SotoCrespo:16,Tikan:18}. Surprisingly, little attention has been paid to the NDR of such universal wave systems up to a very recent theoretical study~\cite{Leisman:EDR:19}. Here, by using an original  strategy, we report the accurate measurement of NDR of  the slowly varying envelop of the waves in one-dimensional deep water waves experiments. We characterize precisely the frequency shift and the broadening of the NDR, which interestingly reveals the presence of solitons and of high order effects. Our results highlight the relevance of the NDR in the context of \textit{integrable turbulence}~\cite{Zakharov:09, Walczak:15,Agafontsev:15, SotoCrespo:16,Tikan:18}.}\\

The dispersion relation plays a key role in wave turbulence (WT) phenomena emerging in the propagation of nonlinear random waves in dispersive media.  In general, WT is described by the resonant interactions among the Fourier components of the wave field~\cite{Zakharov:kolmogorovBook:12}. For example, in a unidirectional wave system dominated by a third-order nonlinearity, the resonance conditions read $\omega_1+\omega_2=\omega_3+\omega_4$ and $k_1+k_2=k_3+k_4$, where $k_i=k(\omega_i)$ satisfies the {\it linear} dispersion relation $k(\omega)$. In this framework, the NDR can be simply defined as  the space-time double Fourier transform of turbulent fields~\cite{Cobelli:DispersionRelation:09, Herbert2010, Aubourg:16}.  The nonlinearity-induced shift and broadening of the NDR  is a fundamental signature of wave turbulence phenomena which has been extensively considered in previous works~\cite{Cobelli:DispersionRelation:09, Herbert2010, Berhanu13, Aubourg:16, Hassaini:Transition:17}.

In comparison with the standard WT, integrable turbulence is of profoundly different nature because the non-trivial resonances are forbidden (see Refs \cite{Zakharov:09,  Agafontsev:15,Walczak:15, SotoCrespo:16,Tikan:18} and Methods). The propagation of solitons is one of the most remarkable properties of integrable systems. Several fundamental questions are still opened in the field of integrable turbulence: what is the mechanism of emergence of localized structures or rogue waves embedded in a random field ? What is the contribution of solitons to the statistical properties ? How does occur the transition between weakly nonlinear random waves and soliton gas~\cite{Hassaini:Transition:17, Redor:19, Suret:SG:20} ? 
Characterizing and understanding the NDR in integrable turbulence is a key step to address these questions. 

In their extensive and very interesting theoretical and numerical work~\cite{Leisman:EDR:19}, Leisman {\it et al.} have recently addressed this question in 1DNLSE for various kinds of initial conditions.

For weak nonlinearity and random waves, the NDR undergoes the well-known frequency shift  called ``Stokes shift'' in the context of water waves~\cite{Stokes1847, Yuen1982} while the NDR of a single soliton is a straight line having a slope corresponding to its group velocity~\cite{Leisman:EDR:19}. It is noteworthy that, to the best of our knowledge, up to now, this spectral signature of solitons has not been reported in the context of integrable turbulence (nonlinear random waves) experiments described by 1DNLSE. \\

In this letter, we examine experimentally and numerically the characteristic features of the NDR developing in integrable turbulence. More precisely, we consider partially coherent (random) waves  initially composed of numerous independent Fourier components (see Methods) and propagating in a unidirectional water tank.\\

We first performed numerical simulations of the 1DNLSE describing the evolution of the slowly varying envelope $\psi(t,z)$ of uni-directional deep water waves (see Methods) for three different  degrees of nonlinearity measured by the parameter $\Gamma$:
\begin{equation} 
\Gamma=\frac{z_{lin}}{z_{nlin}}=\frac{\gamma g P_0}{(2 \pi \Delta f)^2}\;\;\text{with  } P_0=\langle |\psi(t,z=0)|^2 \rangle
\label{eq:Gamma}
\end{equation}
where $\langle ... \rangle$ is the averaging over time and/or realizations, $\gamma$ is the third order nonlinear coefficient, $g$ is the gravity acceleration and $\Delta f$ is the initial spectral width at $z=0$ (see Methods). Fig.~1.a, represents the typical spatio-temporal dynamics of integrable turbulence developing from partially coherent waves in the focusing regime of 1DNLSE. Remarkably, solitons emerge from the turbulent field when the nonlinearity increases.

\begin{figure*}[!t]
  \includegraphics[width=0.75\linewidth]{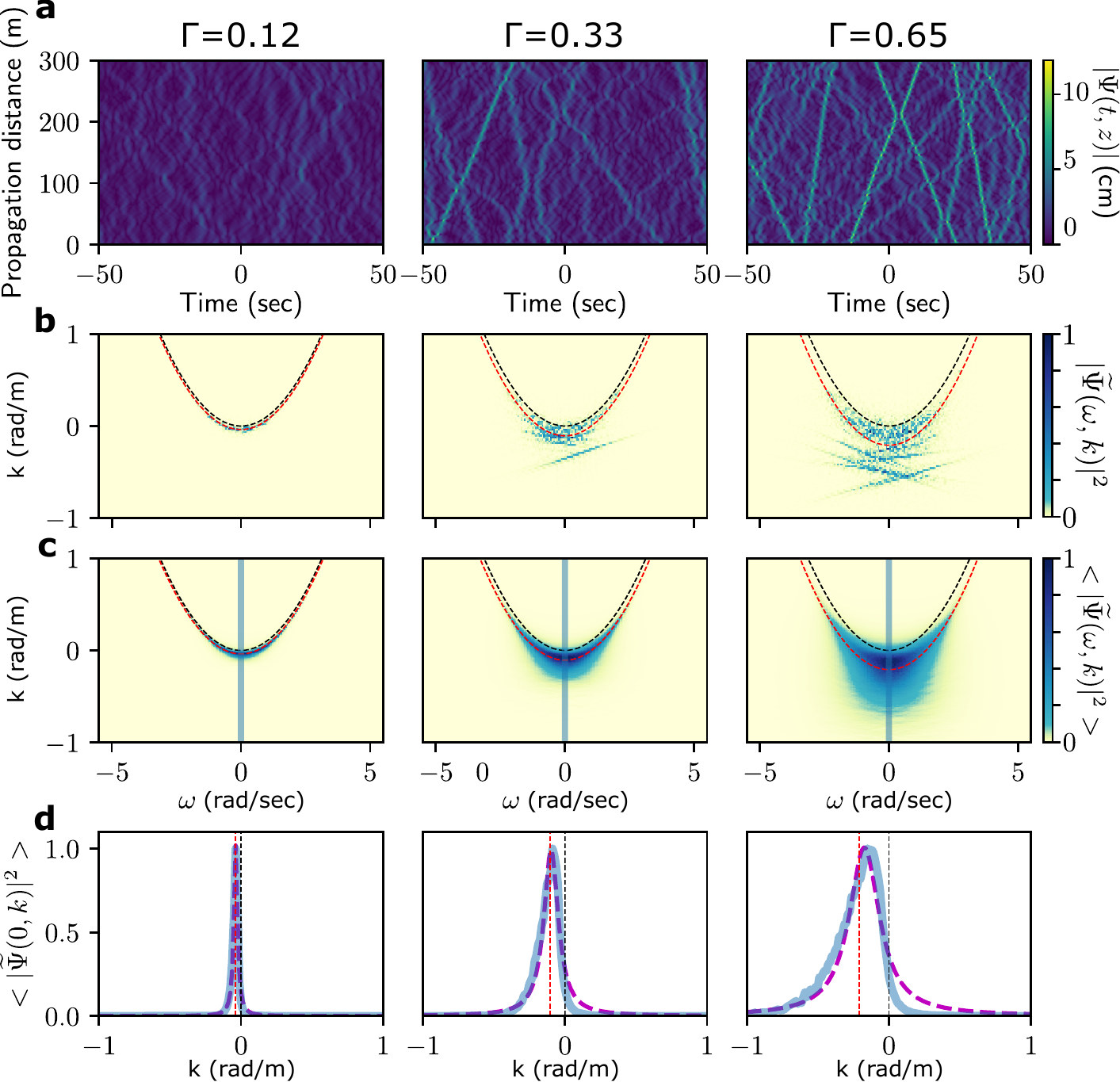}
  \caption{\textbf{Numerical simulations of the 1-D NLS equation}. Three columns correspond to three different values of $\Gamma$ 0.12, 0.33, and 0.65, respectively. The central frequency of the carrier wave and the initial width of the wave spectrum are set to $f_0=1.15$ Hz and $\Delta f = 0.2$ Hz which corresponds to values used in experiments. \textbf{(a)} Spatiotemporal diagram for the complex envelope amplitude $|\psi(t,z)|$. 
  \textbf{(b)} Corresponding nonlinear dispersion relation $|\widetilde{\psi}(\omega,k)|^2$ normalized to the maximum. Black and red curves represent linear dispersion relation and its nonlinear correction according to the expression (\ref{eq:NLDR}). \textbf{(c)} Nonlinear dispersion relation averaged over 1000 realizations and normalized to the maximum. Propagation distance is 500 m. \textbf{(d)} Cross-section of the averaged nonlinear dispersion relation at $\omega=0$ (along the blue line in (c)). Dashed purple line shows a Lorentzian fit. Black and red dashed lines represent the linear dispersion Eq. (\ref{eq:kL})  and its nonlinear correction Eq. (\ref{eq:NLDR}).}
 \label{fig1}
\end{figure*}

\begin{figure*}[!t]
  \includegraphics[width=0.8\linewidth]{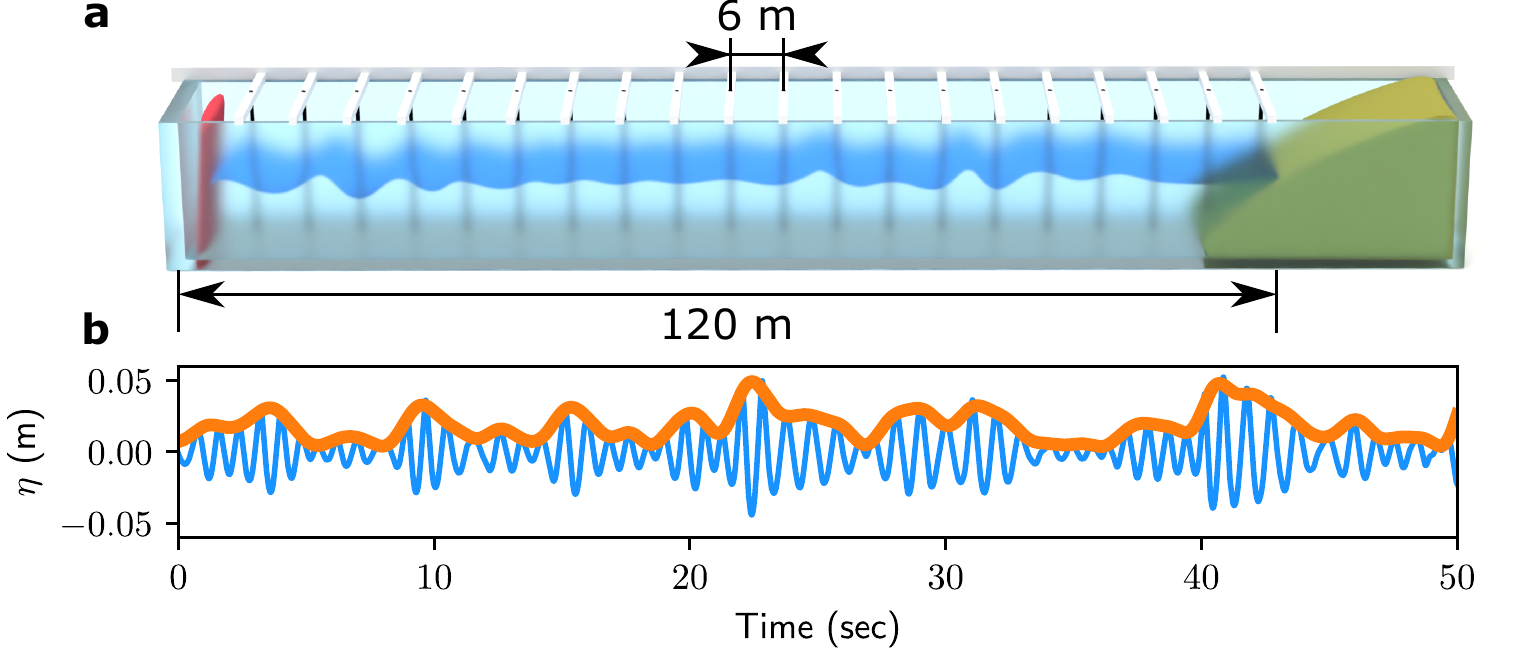}
  \caption{\textbf{Experimental facility.} \textbf{(a)} Schematic representation of the 120m-long water tank facility at \'Ecole  Centrale de Nantes. The surface elevation is recorded by a set of probes equidistantly-placed every 6~m of the water tank length. The water tank is equipped by a parabolically-shaped absorbing beach ($\approx$8~m long) with the addition of pool lanes which provides low back reflection. \textbf{(b)} Typical experimental wave train (surface elevation $\eta$, blue line) and its envelope (orange line) reconstructed by using the Hilbert transform.}
  \label{fig2}
\end{figure*}

\begin{figure*}[!t]
  \includegraphics[width=0.9\linewidth]{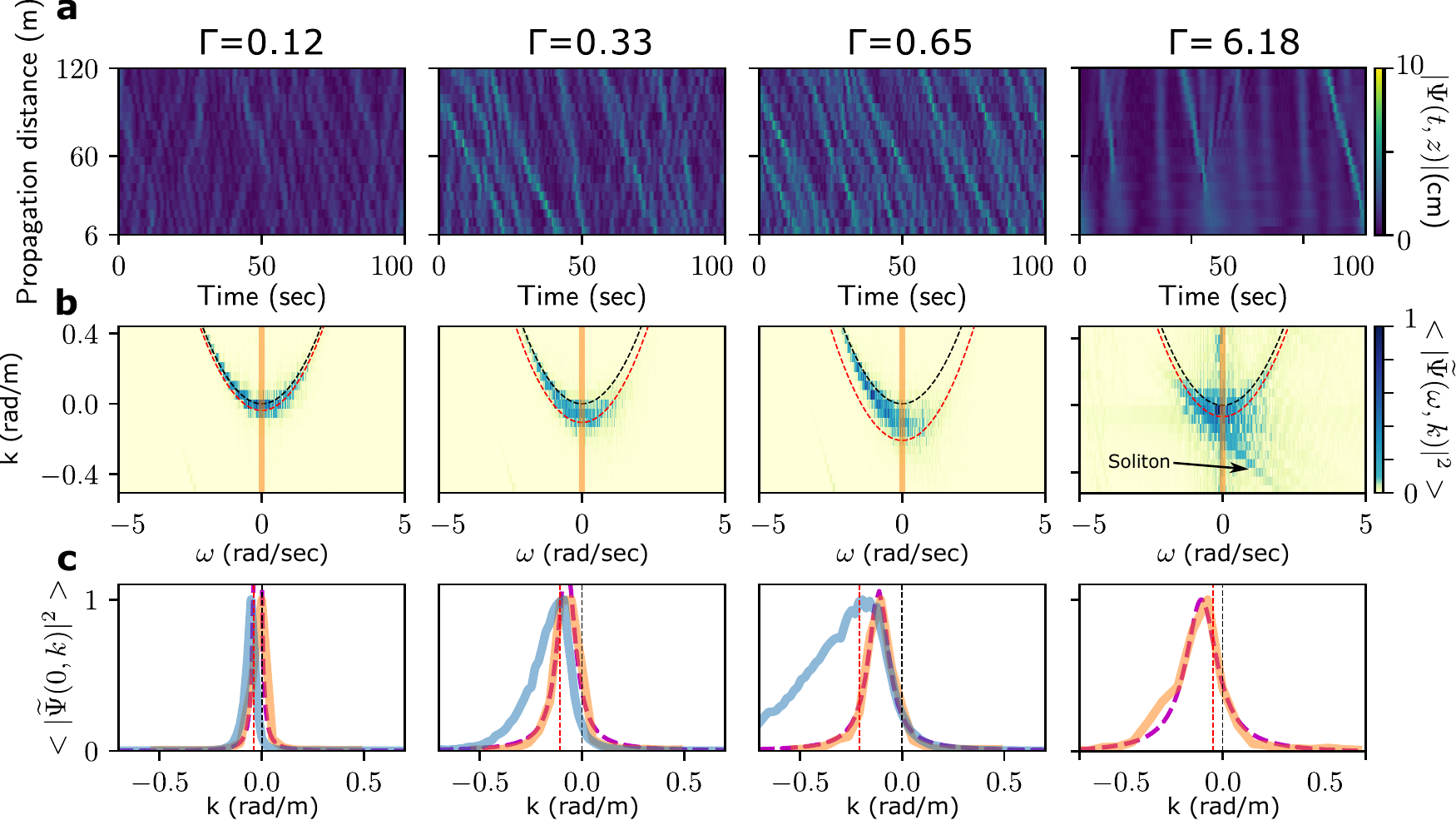}
  \caption{\textbf{Experimental reconstruction of the nonlinear dispersion relation.} Four columns correspond to four different values of $\Gamma$ 0.12, 0.33, 0.65, and 6.18, respectively. \textbf{(a)}  Spatiotemporal diagram of the wave envelope amplitude $|\Psi(t,z)|$. Data received from 20 probes have been post-processed and arranged in 20 vertical rows subtracting waves' group velocity. \textbf{(b)} Nonlinear dispersion relation reconstructed from the evolution of the complex wave envelope. $\Gamma=$ 0.12, 0.33, 0.65 correspond to an initial spectral width $\Delta f=0.2$~Hz and $\langle |\widetilde{\psi}(\omega,k)|^2\rangle$ is averaged over several realization (see Methods). In order to observe the signature of a single soliton, the NDR is not averaged for $\Gamma=6.18$ (corresponding to $\Delta f=0.037$~Hz).   \textbf{(c)} Cross-section of the nonlinear dispersion relation at $\omega=0$. Dashed purple line shows a Lorentzian fit, blue line shows corresponding results of NLS simulation.
  }
\label{fig4}
\end{figure*}

The space-time double Fourier spectrum $|\widetilde{\psi}(\omega,k)|^2$ of {\it one} realization of $\psi(z,t)$  is plotted in blue in Fig.~1.b. As pointed out in~\cite{Leisman:EDR:19}, solitons observed in Fig.~1.a correspond to straight lines in the $k-\omega$ space and  the number of such lines increases when the value of $\Gamma$ is increased. Performing a statistical, we also plot the  spectrum $\langle |\widetilde{\psi}(\omega,k)|^2\rangle$ averaged over several realizations in Fig.~1.c. The linear dispersion of 1DNLSE reads $k(\omega)=\omega^2/g$~\cite{Longuet1962} and is plotted in dashed black lines in Figs 1.b and 1.c. We first evaluate numerically the well-known (Stokes) shift of the dispersion relation induced by nonlinearity. For weak nonlinearities and narrow Fourier spectra made of numerous components, the theoretical nonlinear dispersion relation of 1DNLSE  reads~\cite{Longuet1962, Leisman:EDR:19}:
\begin{equation}
\tilde{k}(\omega,P_0)=k(\omega)-2 \gamma P_0.
\label{eq:NLDR}
\end{equation}

Eq. (\ref{eq:NLDR}) is plotted in red dashed line in Figs~1.b and 1.c. Qualitatively, the shift of the NDR computed from numerical simulations of 1DNLSE follows this theoretical prediction. Moreover, one notices the broadening of the NDR when $\Gamma$ increases. This phenomenon is induced by the energy exchange among Fourier modes and is  well-known in standard WT with resonant interactions~\cite{Cobelli:DispersionRelation:09, Herbert2010, Aubourg:16}. Here, it is remarkable to observe similar behaviour in integrable turbulence in the absence of resonance.

The concept of {\it effective dispersion relation} investigated in the theoretical study~\cite{Lee2009,Leisman:EDR:19} corresponds to the line joining the maxima of the NDR evaluated for each value of $\omega$. In order to quantify not only the shift of this maximum but also the broadening of the NLDR, we plot the $k-$spectrum at $\omega=0$, {\it i.e.} $\langle |\widetilde{\psi}(0,k)|^2\rangle$ in Fig.~1.d. 

For small values of $\Gamma$, the maximum of this curve coincides with the value predicted by the weakly nonlinear theory (Eq.~(\ref{eq:NLDR}) , red dashed line in Fig~1.d).  For this weak nonlinearity regime, note that, at $\omega=0$, the NDR can be empirically fitted by a Lorentzian distribution (purple dashed line in Fig.~1.d). To the best of our knowledge, this remarkable fact is not known for integrable turbulence and has not been  described at the theoretical level. At higher nonlinearities, the shift of the maximum of the NDR is slightly smaller than predicted by the weak nonlinearity theory. Moreover,  the  broadening of the NDR  toward the low values of $k$ observed in Fig~1.c becomes then asymmetric as it is strongly influenced by the emergence of solitons, individually observed in Figs.~1.a and 1.b. ($\Gamma=0.65$).\\

In order to investigate experimentally the NDR described above, we have used the setup described in \cite{Bonnefoy:20}. Unidirectional waves are generated at one end of a $148$ m long, $5$ m wide and $3$ m deep wave flume by using  a computer-assisted flap-type wavemaker (see Fig. 2.a). The flume is equipped with an absorbing device strongly reducing wave reflection at the opposite end. The surface elevation $\eta$ is measured by using $20$ equally spaced resistive wave gauges that are installed along the water tank at distances $z_j=6\,j$ m, $j=1,2,...20$ from the wavemaker located at $z=0$ m. This provides an effective measuring range of $120$~m  and a resolution of the $k-$spectrum  of $2\pi/120$~rad~m$^{-1}$ (see Methods). The envelope $\psi(t,z=0)$ of the surface elevation having a central frequency $f_0=1.15$~Hz is  designed with the same procedure as the one used in our numerical simulations. The degree of nonlinearity is varied by changing the averaged amplitudes or the initial spectral width $\Delta f$ of the waves generated in the water tank.

A typical temporal evolution of the surface elevation experimentally recorded at the first gauge ($z=6$~m) is plotted in the Fig.~2.b.  The slowly varying amplitude $\psi(z,t)$ is determined by using Hilbert Transform. Typical spatio-temporal evolution of $|\psi|$ is plotted in Fig.~3.a where the wave evolution is shown in a retarded frame moving at the group velocity of the carrier wave. When $\Gamma$ increases, the structures become narrower and most of them achieve a negative speed in the $(t,z)$ diagram. 

Accordingly to the numerical simulations reported above, the experimental NDR broadens around the known theoretical shifted parabola given by Eq.~\ref{eq:NLDR} (see Figs. 3.b and 3.c). As expected,  when the degree of nonlinearity increases, the measured NDR broadens and shifts toward the negative values of $k$. Moreover, it becomes asymetric with $\omega$ and deviates significantly from the numerical simulations. This asymetry is a signature in the  $(k-\omega)$ space of the negative speed of the coherent structures in the $(t,z)$ diagram. This phenomenon is the well-known ``frequency downshift'' of surface gravity waves induced by high order nonlinearities~\cite{Trulsen:99}. The High-order spectral (HOS) simulations of our experiments confirm that this shape of the NDR is induced by high order effects (see Methods and Supplementary). 

Moreover,  we found that for $\Gamma \le 1$, straight lines in the ($k-\omega$) space -signatures of solitons- appear much less frequently than in 1DNLSE simulations. However, we have observed this remarkable soliton spectra for small values of the initial spectral width $\Delta f$ corresponding to extremely high nonlinearity (see the fourth column of Fig~3).\\

We now show that the quantitative analysis of the shift and of the broadening of the NDR also reveals the central role played by solitons in (quasi-) integrable turbulence. In Fig. 4,  we plot the maximum $k_M(\Gamma)$ and the full width at half maximum $\Delta k(\Gamma)$  of the NDR $\langle|\widetilde{\psi}(0,k)|^2\rangle$  measured in experiments and computed from numerical simulations of full Euler equations (HOS) and 1DNLSE. The nonlinear phase shift acquired by solitons during their propagation places the solitonic lines well below the dispersion parabola Eq. (\ref{eq:NLDR}) (see Fig~1.b).  The NDR broadens thus asymmetrically and strongly toward negative values of $k$ (see Fig~1.c) while the position of the maximum of the NDR  seems almost to saturate (see Fig.~4.a). As a consequence, for large values of $\Gamma$ ({\it i.e.} high nonlinearity),  $k_M(\Gamma)$ measured in experiments and in simulations evolve more slowly than  predicted by the weakly nonlinear theory - see Eq.~(\ref{eq:NLDR}).

Moreover, Fig. 4 also reveals the soliton inhibition induced by high order effects. Indeed, the shift of $k_M$ toward negative values of $k$ in experiments and HOS simulations is lower than the one predicted by 1DNLSE simulations while the broadening of the NDR is significantly higher in the latter case. Note finally that, remarkably, the shapeq of the experimental and of HOS NDRs coincide with a Lorentzian function for all values of $\Gamma$ while the NDR predicted from the 1DNLSE becomes asymmetric at high nonlinearity (see Fig. 3.c). \\

\begin{figure}[!h]
  \includegraphics[width=1\linewidth]{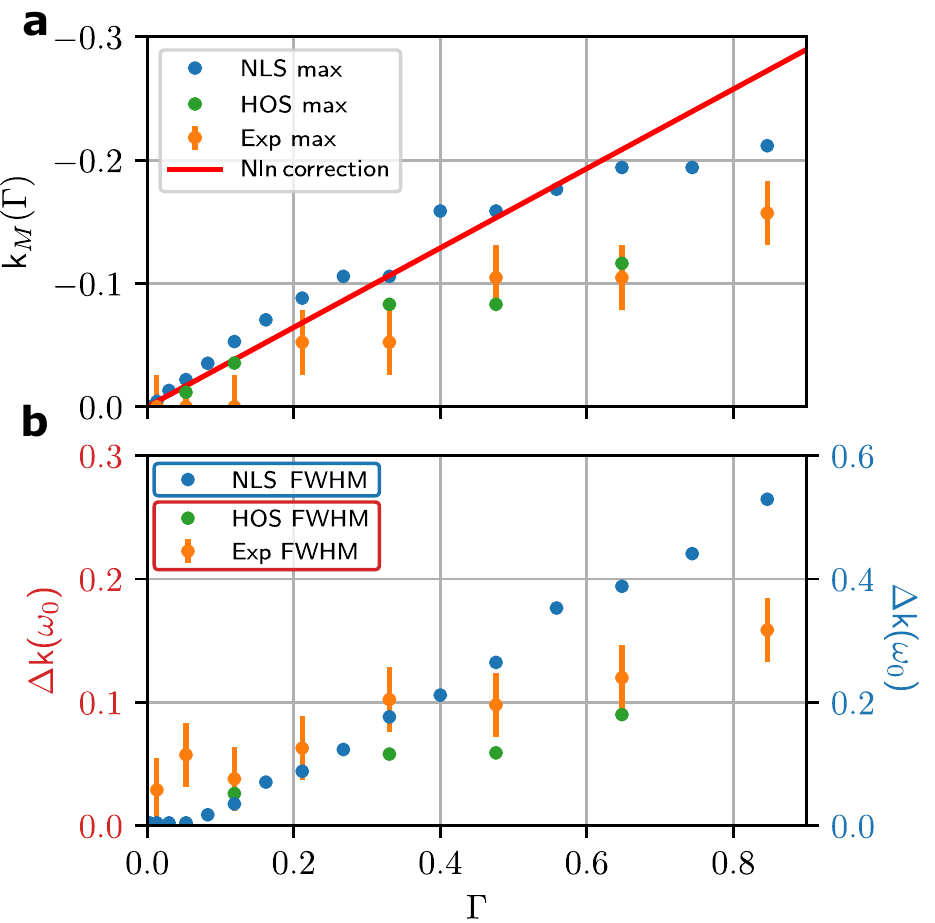}  \caption{\textbf{Quantitative comparison of the experimental results with different numerical models.}  \textbf{(a)} Position of the NDR maximum at central frequency as a function of the parameter $\Gamma$. Blue and green dots correspond to NLS and HOS simulations while orange dots show experimental data. Red line represents the theoretical curve eq.~(\ref{eq:NLDR}). \textbf{(b)} Full width at half maximum of the NDR at zero frequency as a function of the parameter $\Gamma$. Note that the vertical scale is different for experiments/HOS simulations (left scale) and 1DNLSE simulations (right scale). For all points in (a) and (b), f$_0$ = 1.15~Hz, $\Delta$f = 0.2~Hz.}
\label{fig5}
\end{figure}

By focusing our analysis on the slowly varying amplitude $\psi$, we were able to measure very accurately slight deviations from the predicted nonlinear dispersion relation. The results reported here provide new insights into an old fundamental problem of hydrodynamics : the measurement of the dispersion relation of random surface gravity waves  (see for example~\cite{Taklo2015} and refs. therein). Various theoretical and experimental works have been devoted to this question,  see {\it e.g.}~\cite{Longuet1962, Whitham1967, Huang1976, Crawford1981, Wang2004, Gibson2007,Leckler2015, Taklo2015, Taklo2017}. The central point of our strategy is to remove the carrier wave frequency in order to retrieve the NDR of the slowly varying envelope. Our study thus enables an accurate measurement of the nonlinear shift of the dispersion curve together with the broadening and the shape of the NDR  by using a very limited number of 20 gauges (see Methods). For high nonlinearity, our measurements reveal the emergence of solitons embedded in the turbulent field.

Up to now, integrable turbulence has been mainly investigated in optical fibers~\cite{Tikan:18,Walczak:15,Randoux:16}. This study demonstrates that, uni-directional water tanks represent remarkable complementary platforms to investigate the Physics of integrable turbulence. In particular, it is important to note that it is an extremely challenging and open question to measure the NDR of random waves in optical fibers.

The generation of soliton gases in water waves has been recently demonstrated~\cite{Redor:19, Suret:SG:20}. Our work sheds a new light on the importance of the role of solitons in nonlinear random waves. We hope that our work will trigger new experimental and theoretical investigations of NDR in several fields of Physics. In particular, the approach developed in the context of Fermi-Pasta-Ulam-Tsingou system may be promising to predict theoretically the shape and the broadening of the NDR observed here~\cite{Lvov2018}.

\section*{Methods}

\subsection*{One-dimensional Nonlinear Schr\"odinger equation, variables and parameters}
In the context of uni-directionnal deep water waves, 1DNLSE reads:
 \begin{equation}
i\frac{\partial \psi}{\partial z}= \frac{1}{g}\frac{\partial^2 \psi}{\partial t^2}+\gamma |\psi|^2\psi, \label{eq:NLSE}
\end{equation}
where $z$ is the propagation distance, $t$ is the time measured in the frame co-propagating a the group velocity $c_g=g/(2 \omega_0)$  evaluated at the carrier frequency $f_0=\omega_0/(2\pi)$. In the limit of infinite depth, the third order nonlinear coefficient is:
\begin{equation}
\gamma=k_0^3
\end{equation}
where $k_0=\omega_0^2/g$ is the modulus of the wavevector of carrier wave. The surface elevation $\eta(t,z)$ is related at the leading order to the  slowly varying complex envelope $\psi$ as:
\begin{equation}
\eta(t,z)=\frac{1}{2}\left( \psi(t,z)e^{i (k_0z -\omega_0 t)}+c.c.\right); \label{etaA}
\end{equation}

From (\ref{etaA}) the following relation can be derived :
\begin{equation}
\langle|\psi|^2\rangle=2\langle\eta^2\rangle=2 \sigma^2, \label{amplitudesurface}
\end{equation}
where $\langle...\rangle$ implies averages over time and $\sigma^2$ is the variance of the rapidly oscillating wave field.

In order to design the experiment, it is useful to introduce  a linear and a nonlinear propagation length  as follows:
\begin{equation}
z_{lin}=\frac{g}{\Delta \omega^2}\;\;\ {\rm and}\;\;\;\;
z_{nlin}=\frac{1}{\gamma P_0},
\label{eq:zlznl}
\end{equation}
where $\Delta \omega=2\pi \Delta f$ is a typical spectral bandwidth  and $P_0=\langle|\psi_0|^2\rangle$ is the average value of the envelope square both calculated at $z=0$.

The degree of nonlinearity of the wave propagation is given by the parameter:
\begin{equation} 
\Gamma=\frac{z_{lin}}{z_{nlin}}=\frac{\gamma g P_0}{\Delta \omega^2}.
\end{equation}

In the context of ocean waves, $\Gamma=BFI^2$ where BFI is the Benjamin-Feir Index \cite{Janssen:03,Onorato:01}. It has been shown that initial conditions with a large value of $\Gamma$ will eventually lead to the formation of rogue waves characterized by heavy-tailed probability density function of the surface elevation~\cite{Walczak:15,Koussaifi:18,Suret:16,Tikan:18}.

BFI index can be expressed as follows:
\begin{equation}
BFI = \frac{\epsilon}{(\Delta f /f_0)},    
\end{equation}
where $\epsilon= k_0 \sqrt{2} \sigma$ is the wave steepness, and $\Delta f$ and $f_0$ are average spectral width and central frequency of the initial wave packets.  
\\

\subsection*{Resonances in the 1DNLSE}

The non trivial resonances are forbidden in integrable turbulence~\cite{Suret:11}. We consider the third order nonlinear interaction of four monochromatic waves $i=1,2,3,4$ of pulsation $\omega_i$ and wavenumber $k_i(\omega_i)$ in a unidirectional dispersive media. In this context, the resonances conditions of four wave mixing in the standard wave turbulence read $\omega_1+\omega_2=\omega_3+\omega_4$ and $k_1+k_2=k_3+k_4$ where $k_i=k(\omega_i)$. The linear dispersion relation of deep water waves and of the Eq.~({\ref{eq:NLSE}}) is: 

\begin{equation}
    k(\omega)=\frac{\omega^2}{g}
    \label{eq:kL}
\end{equation}

Exact resonances thus lead to:

\begin{equation}
     \omega_1(\omega_3-\omega_2)=\omega_3(\omega_3-\omega_2)
\end{equation}

{\it i.e.} $\omega_1=\omega_3$ and $\omega_2=\omega_4$ or $\omega_2=\omega_3$ and $\omega_1=\omega_4$.  

Exact resonances of non trivial interactions ($\omega_1\ne\omega_3$ and $\omega_1\ne\omega_4$) are thus forbidden.

\subsection*{Initial conditions in simulations and experiments: Partially coherent waves}

In the numerical simulations and in the experiments, the initial conditions are partially coherent waves~\cite{Randoux:16}. The slow varying amplitude of the initial condition at $z=0$ reads:

\begin{equation}
    \psi(t,z=0)=A_0 \sum_{l=-N/2}^{+N/2} e^{-\frac{1}{2}(f_l/\Delta f)^2} e^{i \,2 \pi\,f_l\,t} e^{i\phi_l}
\end{equation}

where $f_l=l/T_{max}$, $T_{max}$ is the temporal duration of the experiments and $\phi_l$ are independently and randomly distributed over $[0,2\pi]$.

 Note that the statistical characteristics of partially coherent waves are very different than plane waves initially perturbed by noise which is also investigated in \cite{Leisman:EDR:19}. For a comparison between the two cases, refer {\it e.g.} to \cite{Copie2020}.

\subsection*{Numerical Simulations}
Numerical simulations of Eq.\ref{eq:NLSE} are realized using step-adaptive high order Runge-Kutta method. We construct different initial conditions using the random phase approach where a uniformly distributed phase is added to every Fourier component of a Gaussian spectrum with $\Delta f=$ 0.2 Hz. Typical temporal windows used in the numerical simulations correspond to 100 seconds.  Three parameters of simulations depend on the value of the steepness : the number of points $N$, the length of propagation $L_{max}$ and the number of realizations $N_{sample}$. We separate the numerical studies into three ranges of steepness $\epsilon$:

\begin{tabular}{llll}
    & $\;\;\;N\;\;\;$ & $L_{max}(m)$ & $N_{sample}$ \\
  $\;\;\;\;\;\;\;\;\;\;\;\epsilon\le 0.05$ & $\;\;2048\;\;$ & $\;\;\;500$ & $\;10000$ \\
  $0.05<\epsilon\le 0.09$ & $\;\;2048\;\;$ & $\;\;\;500$ & $\;\;\;\;500$ \\
  $0.09<\epsilon\le 0.19$ & $\;\;1024\;\;$ & $\;\;2000$ & $\;\;\;\;\;100$ \\
\end{tabular}\\

In order to reconstruct numerically NDR, we multiply the spatiotemporal diagram by a Super-Gaussian window with power 15 along z direction avoiding thereby undesirable effects related to the Fourier analysis of non-periodic signals.

\subsection*{Resolution of the measurement of $k$}

 The key point in our approach is to remove the carrier wave before computing the NDR. This allows us to reveal the details of the NDR of the slowly varying envelop. Note that we only use 20 gauges in the water tank. They are separated by $6$~m and the maximum measurable wavevector is around $1.05$~m$^{-1}$. As a consequence, contrary to Taklo {\it et al.} who use 384 probes, we do not resolve the wavevectors of the carrier wave $2\pi/\lambda_0\simeq5.3$~m$^{-1}$ and of the harmonics. Our strategy enables the accurate  measurement of the NLDR of the slow varying envelop of the wave by using only 20 probes. This provides an effective measuring range of $120$~m with a resolution of the measurement of the $k-$spectrum  of $\Delta k_{min}=2\pi/120$~rad~m$^{-1}$. Note finally that, in~\cite{Taklo2015}, the accuracy of the  measurement of $k$ given by the length of the water tank is $\Delta k_{min}/k_0=0.014$ while our setup enables an accuracy of $\Delta k_{min}/k_0 < 0.01$. 

In order to measure the averaged spectra and NDR, we use 3,6 and, 3  experimental runs with a duration of 512~s for $\Gamma=0.12$, $0.33$, $0.65$, respectively. One run of 128~s have been used for $\Gamma=6.18$.

Note that in NLS and HOS simulations, the chosen lengths of propagation depend on the parameters and vary typically from 300 to 500~m. The uncertainty of measurement of $k_M$ and $\Delta k$ is therefore significantly lower in simulations than in experiments.

\subsection*{Evaluation of the full width at half maximum $\Delta k$ of the position of the maximum $k_M$}

In Fig. 4, we report the evaluation of the full width at half maximum $\Delta k$ and of the position of the maximum $k_M$ of the function $f(k)=|\widetilde{\psi}(k,\omega=0)|^2$ in 1DNLSE, HOS simulations and in experiments. The accuracy of the measurement of $\Delta k$ and $k_M$ is limited both by  the discretization of $k$ (see above the uncertainty $\Delta k_{min}$) and by the random fluctuations of $f(k)$. In order to overcome these difficulties, when it is appropriate, we evaluate $\Delta k$ and $k_M$ by using best fitting procedure with Lorentzian function.

\newpage
\section*{Supplementary Material}

The purpose of this Supplementary Information is to provide some mathematical and numerical details on the High-Order Spectral (HOS) method that is utilized as a reference in the manuscript. Simulation results (spatio-temporal diagrams and NDR) obtained with the model are also presented.

\subsection*{High-Order Spectral method}

One-dimensional Nonlinear Schr\"odinger equation (Eq. (3) of manuscript) is established under the assumptions of weak nonlinearity of the wave field as well as the narrowbandedness of its energy content. Different degrees of nonlinearity ($\Gamma$ parameter) have been tested during the wave tank experiments reported in the manuscript. Then, it appears of great interest to have a high-fidelity numerical model able to replicate the experiments.

This is achieved thanks to the direct numerical simulation of the Euler's equation solved using the High-Order Spectral (HOS) method \cite{dommermuth1987high,west1987new}. The present numerical results have been obtained with the open-source solver HOS-NWT \cite{HOS-NWT}. This model solves the spatio-temporal evolution of the free surface elevation $\eta(z,t)$ in a so-called numerical wave tank. It reproduces all the physical features of the experimental (physical) facility including i) the generation of waves thanks to a wave maker and ii) the absorption of those waves when they reach the opposite wall. More details on the HOS-NWT numerical model as well as different validations performed against wave tank experiments can be found in \cite{bonnefoy2010time,ducrozet2012modified}.

This digital twin is utilized as a complement to the experiments. In the present study, the main advantage of the numerical solution is that it allows to overcome the limitation associated with the finite extent of the physical wave tank (and the associated measuring range $L_m=120~m$). As long as the spatial discretization is kept identical, the numerical wave tank can be of arbitrary size. Increasing the size results in a more accurate resolution $\Delta k_{min}$ for the estimation of wave number $k$, essential to the evaluation of the NDR and its properties (Fig.  4 of the manuscript). \\

\begin{figure*}[!t]
  \includegraphics[width=0.73\linewidth]{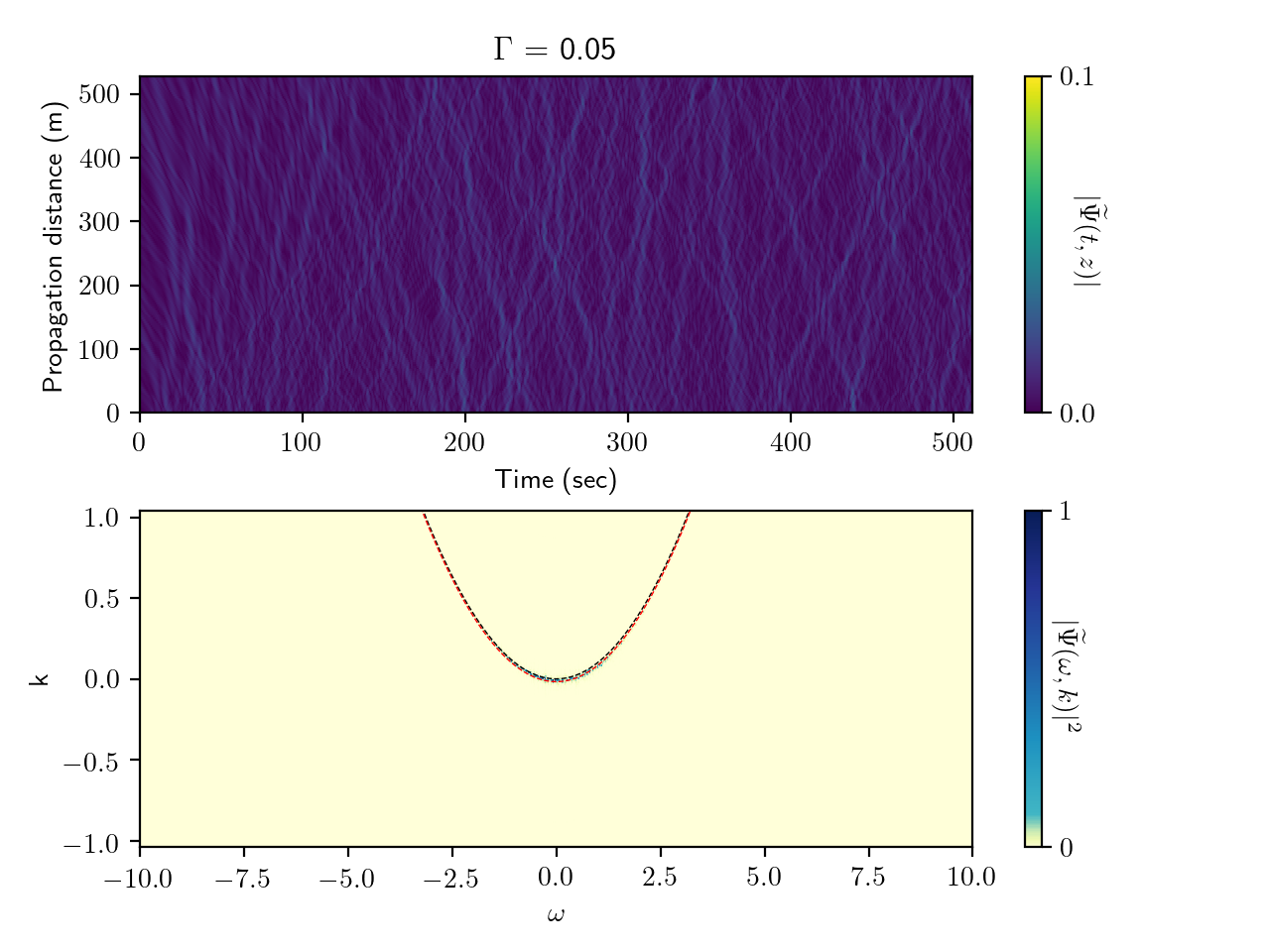}
  \caption{\textbf{Numerical simulation using high-order spectral methods $\Gamma$ = 0.05,
 $\epsilon$ = 0.04.} \textbf{(top)}  Spatiotemporal diagram of the wave envelope amplitude $|\Psi(t,z)|$ evolution over the 531 m of the numerical water tank having 177 probes. \textbf{(bottom)} Nonlinear dispersion relation $|\widetilde{\psi}(\omega,k)|^2$ reconstructed from the evolution of the complex wave envelope.}
  \label{fig1si}
\end{figure*}

\subsection*{Numerical Simulations}

In a wave tank environment, numerical and physical experiments start from temporal initial conditions $\eta(z,t=0)$ at rest. Waves are generated thanks to a wave maker that imposes the spatial initial condition $\eta(z=0,t)$. The HOS digital twin of the \'Ecole Centrale de Nantes water tank facility (Fig. 2.a of original manuscript) uses the exact same wave maker motions than the experimental ones. This allows the direct deterministic comparison between the experimental and numerical wave probes measurements for validation purposes.

Numerical parameters are chosen after a careful convergence study that ensures the accuracy of the numerical solution. The time integration, achieved thanks to a step-adaptive Runge-Kutta method, is controlled by a tolerance parameter chosen as $10^{-8}$ for the present long-time integration. An HOS order of nonlinearity set to $5$ ensures an accurate numerical solution for all configurations tested. Regarding the spatial discretization, the total length of the numerical domain is set to $L_x=560~m$, discretized with $N_x=12288$ points/modes free of aliasing errors.\\

\begin{figure*}[!h]
  \includegraphics[width=0.73\linewidth]{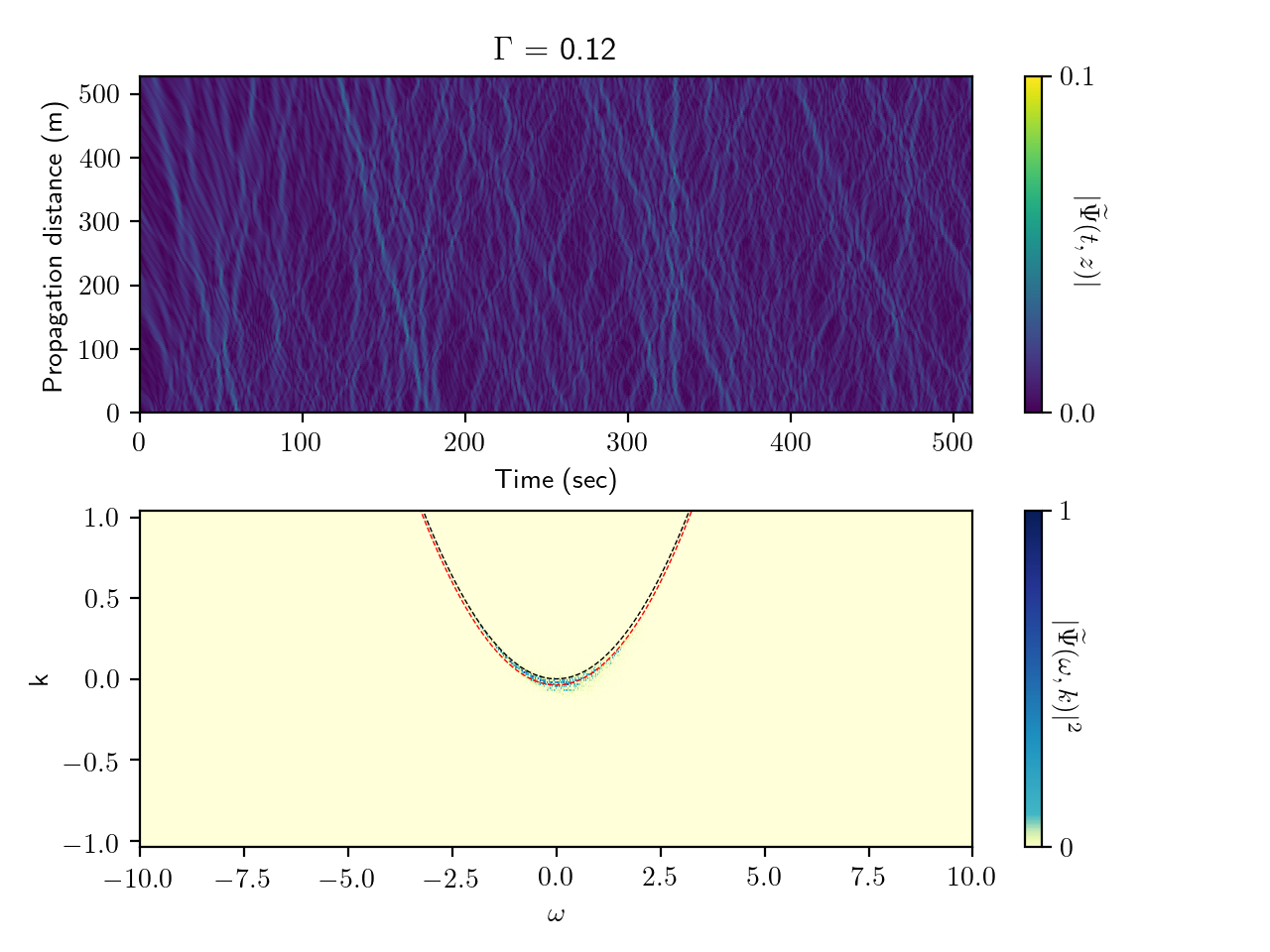}
  \caption{{Numerical simulation using high-order spectral methods $\Gamma$ = 0.12,
 $\epsilon$ = 0.06.} \textbf{(top)}  Spatiotemporal diagram of the wave envelope amplitude $|\Psi(t,z)|$ evolution over the 531 m of the numerical water tank having 177 probes. \textbf{(bottom)} Nonlinear dispersion relation $|\widetilde{\psi}(\omega,k)|^2$ reconstructed from the evolution of the complex wave envelope.}
  \label{fig2si}
\end{figure*}
\begin{figure*}[!h]
  \includegraphics[width=0.73\linewidth]{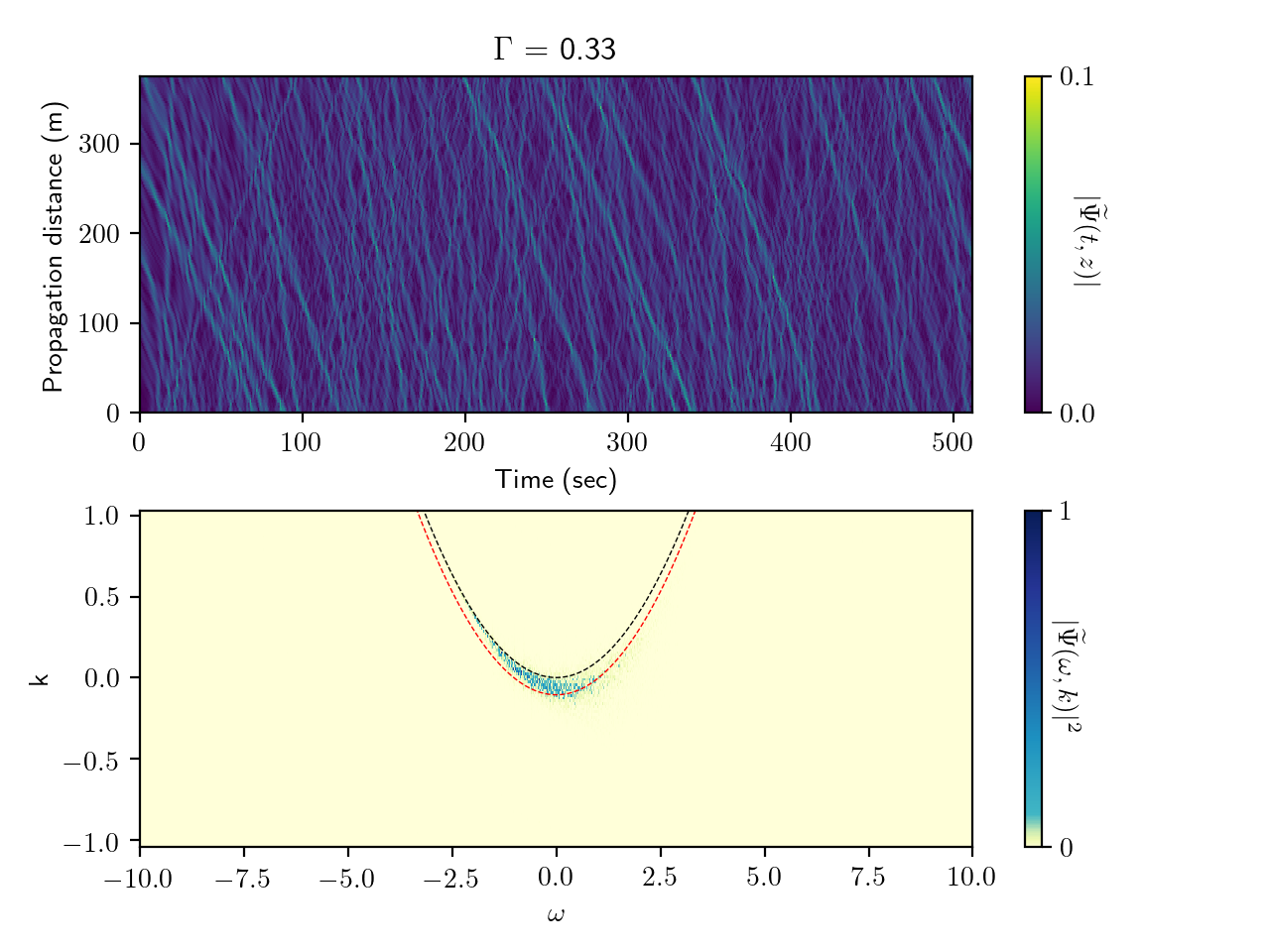}
  \caption{{Numerical simulation using high-order spectral methods $\Gamma$ = 0.33,
 $\epsilon$ = 0.1.} \textbf{(top)}  Spatiotemporal diagram of the wave envelope amplitude $|\Psi(t,z)|$ evolution over the 378 m of the numerical water tank having 126 probes. \textbf{(bottom)} Nonlinear dispersion relation $|\widetilde{\psi}(\omega,k)|^2$ reconstructed from the evolution of the complex wave envelope.}
  \label{fig3si}
\end{figure*}
\begin{figure*}[!h]
  \includegraphics[width=0.73\linewidth]{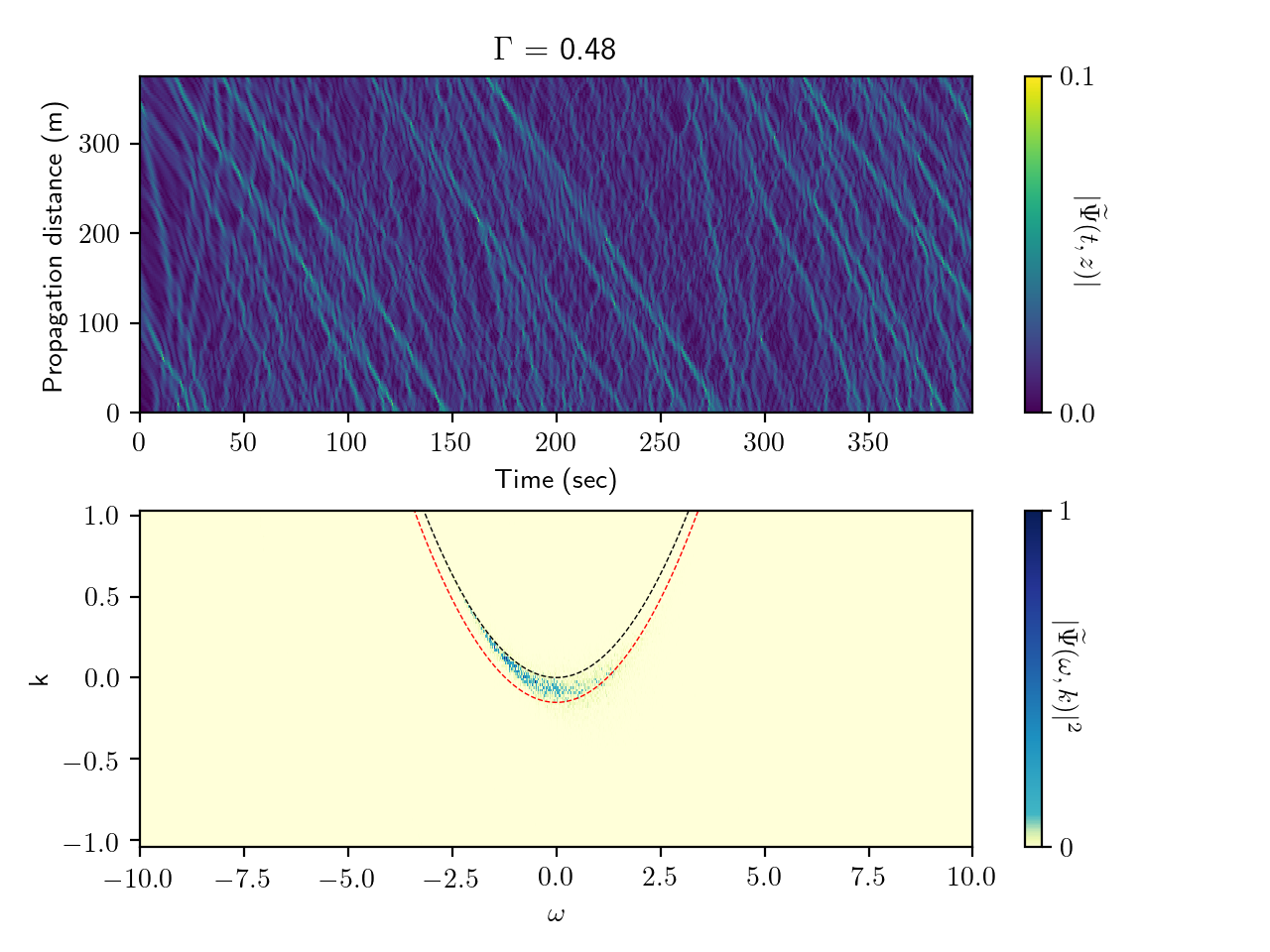}
  \caption{{Numerical simulation using high-order spectral methods $\Gamma$ = 0.48,
 $\epsilon$ = 0.12.} \textbf{(top)}  Spatiotemporal diagram of the wave envelope amplitude $|\Psi(t,z)|$ evolution over the 378 m of the numerical water tank having 126 probes. \textbf{(bottom)} Nonlinear dispersion relation $|\widetilde{\psi}(\omega,k)|^2$ reconstructed from the evolution of the complex wave envelope.}
  \label{fig4si}
\end{figure*}
\begin{figure*}[!h]
  \includegraphics[width=0.73\linewidth]{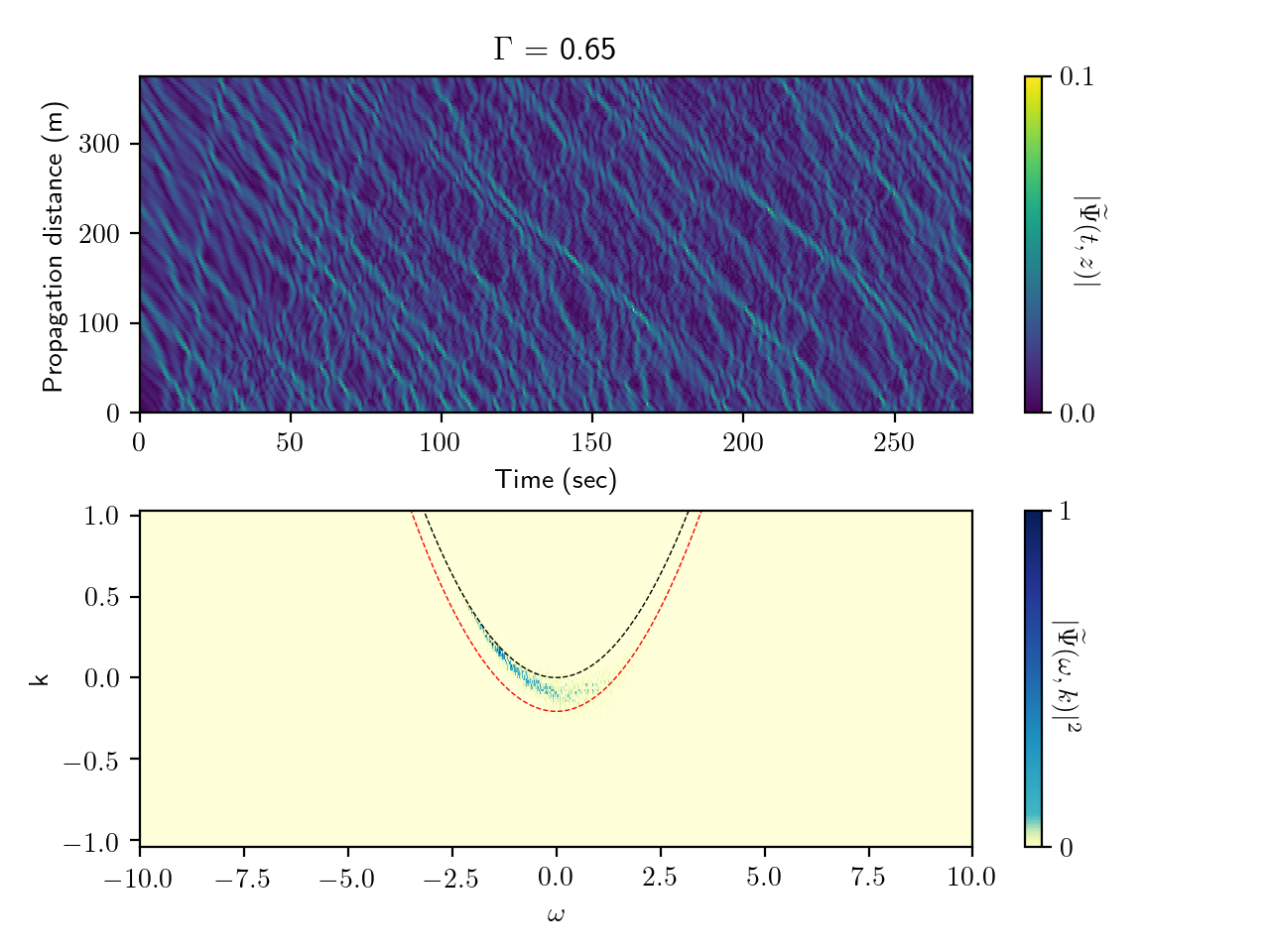}
  \caption{{Numerical simulation using high-order spectral methods $\Gamma$ = 0.65,
 $\epsilon$ = 0.14.} \textbf{(top)}  Spatiotemporal diagram of the wave envelope amplitude $|\Psi(t,z)|$ evolution over the 378 m of the numerical water tank having 126 probes. \textbf{(bottom)} Nonlinear dispersion relation $|\widetilde{\psi}(\omega,k)|^2$ reconstructed from the evolution of the complex wave envelope.}
  \label{fig5si}
\end{figure*}

The wave conditions simulated cover a wide range of nonlinearity with $\Gamma=\left[ 0.05 ; 0.12 ; 0.33 ; 0.48 ; 0.65\right]$. The spatiotemporal diagram of the wave enveloped amplitude $|\Psi(t,z)|$ as well as the corresponding NDR $|\widetilde{\psi}(k,\omega)|^2$ are presented in Figures \ref{fig1si} to \ref{fig5si} with an increasing level of nonlinearity. 

Taking into account the evolution of the nonlinear scale $z_{nlin}$ with $\Gamma$ (Eqs. (7) and (8) of manuscript), the analysis is performed on a length of $531~m$ for $\Gamma=0.05$ and $\Gamma=0.12$ in Figs. \ref{fig1si} and \ref{fig2si} respectively. With an increased nonlinearity $\Gamma=\left[ 0.33 ; 0.48 ; 0.65\right]$, the analysis is conducted on a shorter length of $378~m$. Similarly to the experimental facility, the numerical results use wave gauges that are equally spaced in the computational domain every $6~m$.\\

Overall, the HOS numerical simulations confirm observations in the physical wave tank. As expected, comparing left column of Fig. 1 in the manuscript and Fig. SI \ref{fig2si}, for small values of $\Gamma$ 1DNLSE and HOS results are very similar with a small departure from linear dispersion relation. Then, an increased nonlinearity is associated with coherent structures that are more visible in the $(t,z)$ diagram, see Fig. SI \ref{fig5si} for instance. However, compared to the 1DNLSE results (Fig.1 right column of manuscript), the corresponding straight lines are less frequent in the Euler's simulations. In addition, the evolution of those structures is clearly subjected to a negative speed when taking into account high-order nonlinear and dispersion effects. The corresponding frequency down-shift is also clearly visible in the NDR.

 \clearpage 

\begin{acknowledgments}
This work has been partially supported  by the Agence Nationale de la
Recherche  through the LABEX CEMPI project (ANR-11-LABX-0007) and the ANR DYSTURB Project (ANR-17-CE30-0004),   the
Ministry of Higher Education and Research, Hauts de France council and
European Regional Development  Fund (ERDF) through the the Nord-Pas de
Calais Regional Research Council and the European Regional Development
Fund (ERDF) through the Contrat de Projets Etat-R\'egion (CPER
Photonics for Society P4S).  E.F. is thankful for partial support from the Simons Foundation/MPS No. 651463
\end{acknowledgments}


\begin{thebibliography}{10}
\expandafter\ifx\csname url\endcsname\relax
  \def\url#1{\texttt{#1}}\fi
\expandafter\ifx\csname urlprefix\endcsname\relax\def\urlprefix{URL }\fi
\providecommand{\bibinfo}[2]{#2}
\providecommand{\eprint}[2][]{\url{#2}}

\bibitem{Hutchings1992}
\bibinfo{author}{Hutchings, D.~C.}, \bibinfo{author}{Sheik-Bahae, M.},
  \bibinfo{author}{Hagan, D.~J.} \& \bibinfo{author}{Van~Stryland, E.~W.}
\newblock \bibinfo{title}{Kramers-kr{\"o}nig relations in nonlinear optics}.
\newblock \emph{\bibinfo{journal}{Optical and Quantum Electronics}}
  \textbf{\bibinfo{volume}{24}}, \bibinfo{pages}{1--30} (\bibinfo{year}{1992}).

\bibitem{Herbert2010}
\bibinfo{author}{Herbert, E.}, \bibinfo{author}{Mordant, N.} \&
  \bibinfo{author}{Falcon, E.}
\newblock \bibinfo{title}{Observation of the nonlinear dispersion relation and
  spatial statistics of wave turbulence on the surface of a fluid}.
\newblock \emph{\bibinfo{journal}{Physical review letters}}
  \textbf{\bibinfo{volume}{105}}, \bibinfo{pages}{144502}
  (\bibinfo{year}{2010}).

\bibitem{Taklo2015}
\bibinfo{author}{Taklo, T. M.~A.}, \bibinfo{author}{Trulsen, K.},
  \bibinfo{author}{Gramstad, O.}, \bibinfo{author}{Krogstad, H.~E.} \&
  \bibinfo{author}{Jensen, A.}
\newblock \bibinfo{title}{Measurement of the dispersion relation for random
  surface gravity waves}.
\newblock \emph{\bibinfo{journal}{Journal of Fluid Mechanics}}
  (\bibinfo{year}{2015}).

\bibitem{Aubourg:16}
\bibinfo{author}{Aubourg, Q.} \& \bibinfo{author}{Mordant, N.}
\newblock \bibinfo{title}{Investigation of resonances in gravity-capillary wave
  turbulence}.
\newblock \emph{\bibinfo{journal}{Physical Review Fluids}}
  \textbf{\bibinfo{volume}{1}}, \bibinfo{pages}{023701} (\bibinfo{year}{2016}).

\bibitem{deike13}
\bibinfo{author}{Deike, L.}, \bibinfo{author}{Bacri, J.-C.} \&
  \bibinfo{author}{Falcon, E.}
\newblock \bibinfo{title}{Nonlinear waves on the surface of a fluid covered by
  an elastic sheet}.
\newblock \emph{\bibinfo{journal}{Journal of Fluid Mechanics}}
  \textbf{\bibinfo{volume}{733}}, \bibinfo{pages}{394–413}
  (\bibinfo{year}{2013}).

\bibitem{Cobelli:DispersionRelation:09}
\bibinfo{author}{Cobelli, P.}, \bibinfo{author}{Petitjeans, P.},
  \bibinfo{author}{Maurel, A.}, \bibinfo{author}{Pagneux, V.} \&
  \bibinfo{author}{Mordant, N.}
\newblock \bibinfo{title}{Space-time resolved wave turbulence in a vibrating
  plate}.
\newblock \emph{\bibinfo{journal}{Physical review letters}}
  \textbf{\bibinfo{volume}{103}}, \bibinfo{pages}{204301}
  (\bibinfo{year}{2009}).

\bibitem{Carretero2008}
\bibinfo{author}{Carretero-Gonz{\'a}lez, R.}, \bibinfo{author}{Frantzeskakis,
  D.} \& \bibinfo{author}{Kevrekidis, P.}
\newblock \bibinfo{title}{Nonlinear waves in bose--einstein condensates:
  physical relevance and mathematical techniques}.
\newblock \emph{\bibinfo{journal}{Nonlinearity}} \textbf{\bibinfo{volume}{21}},
  \bibinfo{pages}{R139} (\bibinfo{year}{2008}).

\bibitem{Benisti2008}
\bibinfo{author}{Benisti, D.}, \bibinfo{author}{Strozzi, D.~J.} \&
  \bibinfo{author}{Gremillet, L.}
\newblock \bibinfo{title}{Breakdown of electrostatic predictions for the
  nonlinear dispersion relation of a stimulated raman scattering driven plasma
  wave}.
\newblock \emph{\bibinfo{journal}{Physics of Plasmas}}
  \textbf{\bibinfo{volume}{15}}, \bibinfo{pages}{030701}
  (\bibinfo{year}{2008}).

\bibitem{Zakharov:09}
\bibinfo{author}{Zakharov, V.~E.}
\newblock \bibinfo{title}{Turbulence in integrable systems}.
\newblock \emph{\bibinfo{journal}{Stud. Appl. Math.}}
  \textbf{\bibinfo{volume}{122}}, \bibinfo{pages}{219--234}
  (\bibinfo{year}{2009}).

\bibitem{Walczak:15}
\bibinfo{author}{Walczak, P.}, \bibinfo{author}{Randoux, S.} \&
  \bibinfo{author}{Suret, P.}
\newblock \bibinfo{title}{Optical rogue waves in integrable turbulence}.
\newblock \emph{\bibinfo{journal}{Phys. Rev. Lett.}}
  \textbf{\bibinfo{volume}{114}}, \bibinfo{pages}{143903}
  (\bibinfo{year}{2015}).

\bibitem{Agafontsev:15}
\bibinfo{author}{Agafontsev, D.} \& \bibinfo{author}{Zakharov, V.~E.}
\newblock \bibinfo{title}{Integrable turbulence and formation of rogue waves}.
\newblock \emph{\bibinfo{journal}{Nonlinearity}} \textbf{\bibinfo{volume}{28}},
  \bibinfo{pages}{2791} (\bibinfo{year}{2015}).

\bibitem{SotoCrespo:16}
\bibinfo{author}{Soto-Crespo, J.~M.}, \bibinfo{author}{Devine, N.} \&
  \bibinfo{author}{Akhmediev, N.}
\newblock \bibinfo{title}{Integrable turbulence and rogue waves: Breathers or
  solitons?}
\newblock \emph{\bibinfo{journal}{Phys. Rev. Lett.}}
  \textbf{\bibinfo{volume}{116}}, \bibinfo{pages}{103901}
  (\bibinfo{year}{2016}).

\bibitem{Tikan:18}
\bibinfo{author}{Tikan, A.}, \bibinfo{author}{Bielawski, S.},
  \bibinfo{author}{Szwaj, C.}, \bibinfo{author}{Randoux, S.} \&
  \bibinfo{author}{Suret, P.}
\newblock \bibinfo{title}{Single-shot measurement of phase and amplitude by
  using a heterodyne time-lens system and ultrafast digital time-holography}.
\newblock \emph{\bibinfo{journal}{Nature Photonics}}
  \textbf{\bibinfo{volume}{12}}, \bibinfo{pages}{228} (\bibinfo{year}{2018}).

\bibitem{Leisman:EDR:19}
\bibinfo{author}{Leisman, K.~P.}, \bibinfo{author}{Zhou, D.},
  \bibinfo{author}{Banks, J.}, \bibinfo{author}{Kova{\v{c}}i{\v{c}}, G.} \&
  \bibinfo{author}{Cai, D.}
\newblock \bibinfo{title}{Effective dispersion in the focusing nonlinear
  schr{\"o}dinger equation}.
\newblock \emph{\bibinfo{journal}{Physical Review E}}
  \textbf{\bibinfo{volume}{100}}, \bibinfo{pages}{022215}
  (\bibinfo{year}{2019}).

\bibitem{Zakharov:kolmogorovBook:12}
\bibinfo{author}{Zakharov, V.~E.}, \bibinfo{author}{L'vov, V.~S.} \&
  \bibinfo{author}{Falkovich, G.}
\newblock \emph{\bibinfo{title}{Kolmogorov spectra of turbulence I: Wave
  turbulence}} (\bibinfo{publisher}{Springer Science \& Business Media},
  \bibinfo{year}{2012}).

\bibitem{Berhanu13}
\bibinfo{author}{Berhanu, M.} \& \bibinfo{author}{Falcon, E.}
\newblock \bibinfo{title}{Space-time-resolved capillary wave turbulence}.
\newblock \emph{\bibinfo{journal}{Phys. Rev. E}} \textbf{\bibinfo{volume}{87}},
  \bibinfo{pages}{033003} (\bibinfo{year}{2013}).
\newblock \urlprefix\url{https://link.aps.org/doi/10.1103/PhysRevE.87.033003}.

\bibitem{Hassaini:Transition:17}
\bibinfo{author}{Hassaini, R.} \& \bibinfo{author}{Mordant, N.}
\newblock \bibinfo{title}{Transition from weak wave turbulence to soliton gas}.
\newblock \emph{\bibinfo{journal}{Physical Review Fluids}}
  \textbf{\bibinfo{volume}{2}}, \bibinfo{pages}{094803} (\bibinfo{year}{2017}).

\bibitem{Redor:19}
\bibinfo{author}{Redor, I.}, \bibinfo{author}{Barth\'elemy, E.},
  \bibinfo{author}{Michallet, H.}, \bibinfo{author}{Onorato, M.} \&
  \bibinfo{author}{Mordant, N.}
\newblock \bibinfo{title}{Experimental evidence of a hydrodynamic soliton gas}.
\newblock \emph{\bibinfo{journal}{Phys. Rev. Lett.}}
  \textbf{\bibinfo{volume}{122}}, \bibinfo{pages}{214502}
  (\bibinfo{year}{2019}).
\newblock
  \urlprefix\url{https://link.aps.org/doi/10.1103/PhysRevLett.122.214502}.

\bibitem{Suret:SG:20}
\bibinfo{author}{Suret, P.} \emph{et~al.}
\newblock \bibinfo{title}{Nonlinear spectral synthesis of soliton gas in
  deep-water surface gravity waves}.
\newblock \emph{\bibinfo{journal}{Physical Review Letters}}
  \textbf{\bibinfo{volume}{125}}, \bibinfo{pages}{264101}
  (\bibinfo{year}{2020}).

\bibitem{Stokes1847}
\bibinfo{author}{Stokes, G.}
\newblock \bibinfo{title}{On the theory of oscillatory waves}.
\newblock \emph{\bibinfo{journal}{Transactions of the Cambridge Philosophical
  Society}} \textbf{\bibinfo{volume}{8}}, \bibinfo{pages}{441--455}
  (\bibinfo{year}{1847}).

\bibitem{Yuen1982}
\bibinfo{author}{Yuen, H.~C.} \& \bibinfo{author}{Lake, B.~M.}
\newblock \bibinfo{title}{Nonlinear dynamics of deep-water gravity waves}.
\newblock In \emph{\bibinfo{booktitle}{Advances in applied mechanics}},
  vol.~\bibinfo{volume}{22}, \bibinfo{pages}{67--229}
  (\bibinfo{publisher}{Elsevier}, \bibinfo{year}{1982}).

\bibitem{Longuet1962}
\bibinfo{author}{Longuet-Higgins, M.~S.} \& \bibinfo{author}{Phillips, O.~M.}
\newblock \bibinfo{title}{Phase velocity effects in tertiary wave
  interactions}.
\newblock \emph{\bibinfo{journal}{Journal of Fluid Mechanics}}
  \textbf{\bibinfo{volume}{12}}, \bibinfo{pages}{333--336}
  (\bibinfo{year}{1962}).

\bibitem{Lee2009}
\bibinfo{author}{Lee, W.}, \bibinfo{author}{Kova\ifmmode \check{c}\else
  \v{c}\fi{}i\ifmmode~\check{c}\else \v{c}\fi{}, G.} \& \bibinfo{author}{Cai,
  D.}
\newblock \bibinfo{title}{Renormalized resonance quartets in dispersive wave
  turbulence}.
\newblock \emph{\bibinfo{journal}{Phys. Rev. Lett.}}
  \textbf{\bibinfo{volume}{103}}, \bibinfo{pages}{024502}
  (\bibinfo{year}{2009}).
\newblock
  \urlprefix\url{https://link.aps.org/doi/10.1103/PhysRevLett.103.024502}.

\bibitem{Bonnefoy:20}
\bibinfo{author}{Bonnefoy, F.} \emph{et~al.}
\newblock \bibinfo{title}{From modulational instability to focusing dam breaks
  in water waves}.
\newblock \emph{\bibinfo{journal}{Phys. Rev. Fluids}}
  \textbf{\bibinfo{volume}{5}}, \bibinfo{pages}{034802} (\bibinfo{year}{2020}).

\bibitem{Trulsen:99}
\bibinfo{author}{Trulsen, K.}, \bibinfo{author}{Stansberg, C.~T.} \&
  \bibinfo{author}{Velarde, M.~G.}
\newblock \bibinfo{title}{Laboratory evidence of three-dimensional frequency
  downshift of waves in a long tank}.
\newblock \emph{\bibinfo{journal}{Physics of Fluids}}
  \textbf{\bibinfo{volume}{11}}, \bibinfo{pages}{235--237}
  (\bibinfo{year}{1999}).

\bibitem{Whitham1967}
\bibinfo{author}{Whitham, G.}
\newblock \bibinfo{title}{Non-linear dispersion of water waves}.
\newblock \emph{\bibinfo{journal}{Journal of Fluid Mechanics}}
  \textbf{\bibinfo{volume}{27}}, \bibinfo{pages}{399--412}
  (\bibinfo{year}{1967}).

\bibitem{Huang1976}
\bibinfo{author}{Huang, N.~E.} \& \bibinfo{author}{Tung, C.-C.}
\newblock \bibinfo{title}{The dispersion relation for a nonlinear random
  gravity wave field}.
\newblock \emph{\bibinfo{journal}{Journal of Fluid Mechanics}}
  \textbf{\bibinfo{volume}{75}}, \bibinfo{pages}{337--345}
  (\bibinfo{year}{1976}).

\bibitem{Crawford1981}
\bibinfo{author}{Crawford, D.~R.}, \bibinfo{author}{Lake, B.~M.} \&
  \bibinfo{author}{Yuen, H.~C.}
\newblock \bibinfo{title}{Effects of nonlinearity and spectral bandwidth on the
  dispersion relation and component phase speeds of surface gravity waves}.
\newblock \emph{\bibinfo{journal}{Journal of Fluid Mechanics}}
  \textbf{\bibinfo{volume}{112}}, \bibinfo{pages}{1--32}
  (\bibinfo{year}{1981}).

\bibitem{Wang2004}
\bibinfo{author}{Wang, D.~W.} \& \bibinfo{author}{Hwang, P.~A.}
\newblock \bibinfo{title}{The dispersion relation of short wind waves from
  space--time wave measurements}.
\newblock \emph{\bibinfo{journal}{Journal of Atmospheric and Oceanic
  Technology}} \textbf{\bibinfo{volume}{21}}, \bibinfo{pages}{1936--1945}
  (\bibinfo{year}{2004}).

\bibitem{Gibson2007}
\bibinfo{author}{Gibson, R.} \& \bibinfo{author}{Swan, C.}
\newblock \bibinfo{title}{The evolution of large ocean waves: the role of local
  and rapid spectral changes}.
\newblock \emph{\bibinfo{journal}{Proceedings of the Royal Society A:
  Mathematical, Physical and Engineering Sciences}}
  \textbf{\bibinfo{volume}{463}}, \bibinfo{pages}{21--48}
  (\bibinfo{year}{2007}).

\bibitem{Leckler2015}
\bibinfo{author}{Leckler, F.} \emph{et~al.}
\newblock \bibinfo{title}{Analysis and interpretation of frequency--wavenumber
  spectra of young wind waves}.
\newblock \emph{\bibinfo{journal}{Journal of Physical Oceanography}}
  \textbf{\bibinfo{volume}{45}}, \bibinfo{pages}{2484--2496}
  (\bibinfo{year}{2015}).

\bibitem{Taklo2017}
\bibinfo{author}{Taklo, T. M.~A.}, \bibinfo{author}{Trulsen, K.},
  \bibinfo{author}{Krogstad, H.~E.} \& \bibinfo{author}{Borge, J. C.~N.}
\newblock \bibinfo{title}{On dispersion of directional surface gravity waves}.
\newblock \emph{\bibinfo{journal}{Journal Of Fluid Mechanics}}
  \textbf{\bibinfo{volume}{812}}, \bibinfo{pages}{681--697}
  (\bibinfo{year}{2017}).

\bibitem{Randoux:16}
\bibinfo{author}{Randoux, S.}, \bibinfo{author}{Walczak, P.},
  \bibinfo{author}{Onorato, M.} \& \bibinfo{author}{Suret, P.}
\newblock \bibinfo{title}{Nonlinear random optical waves: Integrable
  turbulence, rogue waves and intermittency}.
\newblock \emph{\bibinfo{journal}{Physica D: Nonlinear Phenomena}}
  \bibinfo{pages}{--} (\bibinfo{year}{2016}).
\newblock
  \urlprefix\url{http://www.sciencedirect.com/science/article/pii/S0167278916301506}.

\bibitem{Lvov2018}
\bibinfo{author}{Lvov, Y.~V.} \& \bibinfo{author}{Onorato, M.}
\newblock \bibinfo{title}{Double scaling in the relaxation time in the
  $\beta$-fermi-pasta-ulam-tsingou model}.
\newblock \emph{\bibinfo{journal}{Physical review letters}}
  \textbf{\bibinfo{volume}{120}}, \bibinfo{pages}{144301}
  (\bibinfo{year}{2018}).

\bibitem{Janssen:03}
\bibinfo{author}{Janssen, P. A. E.~M.}
\newblock \bibinfo{title}{Nonlinear four-wave interactions and freak waves}.
\newblock \emph{\bibinfo{journal}{J. Phys. Oceanogr.}}
  \textbf{\bibinfo{volume}{33}}, \bibinfo{pages}{863} (\bibinfo{year}{2003}).

\bibitem{Onorato:01}
\bibinfo{author}{Onorato, M.}, \bibinfo{author}{Osborne, A.~R.},
  \bibinfo{author}{Serio, M.} \& \bibinfo{author}{Bertone, S.}
\newblock \bibinfo{title}{Freak waves in random oceanic sea states}.
\newblock \emph{\bibinfo{journal}{Phys. Rev. Lett.}}
  \textbf{\bibinfo{volume}{86}}, \bibinfo{pages}{5831--5834}
  (\bibinfo{year}{2001}).

\bibitem{Koussaifi:18}
\bibinfo{author}{El~Koussaifi, R.} \emph{et~al.}
\newblock \bibinfo{title}{Spontaneous emergence of rogue waves in partially
  coherent waves: a quantitative experimental comparison between hydrodynamics
  and optics}.
\newblock \emph{\bibinfo{journal}{Physical Review E}}
  \textbf{\bibinfo{volume}{97}}, \bibinfo{pages}{012208}
  (\bibinfo{year}{2018}).

\bibitem{Suret:16}
\bibinfo{author}{Suret, P.} \emph{et~al.}
\newblock \bibinfo{title}{Single-shot observation of optical rogue waves in
  integrable turbulence using time microscopy}.
\newblock \emph{\bibinfo{journal}{Nat. Commun.}} \textbf{\bibinfo{volume}{7}}
  (\bibinfo{year}{2016}).

\bibitem{Suret:11}
\bibinfo{author}{Suret, P.}, \bibinfo{author}{Picozzi, A.} \&
  \bibinfo{author}{Randoux, S.}
\newblock \bibinfo{title}{Wave turbulence in integrable systems: nonlinear
  propagation of incoherent optical waves in single-mode fibers}.
\newblock \emph{\bibinfo{journal}{Opt. Express}} \textbf{\bibinfo{volume}{19}},
  \bibinfo{pages}{17852--17863} (\bibinfo{year}{2011}).

\bibitem{Copie2020}
\bibinfo{author}{Copie, F.}, \bibinfo{author}{Randoux, S.} \&
  \bibinfo{author}{Suret, P.}
\newblock \bibinfo{title}{The physics of the one-dimensional nonlinear
  schr{\"o}dinger equation in fiber optics: rogue waves, modulation instability
  and self-focusing phenomena}.
\newblock \emph{\bibinfo{journal}{Reviews in Physics}}
  \textbf{\bibinfo{volume}{5}}, \bibinfo{pages}{100037} (\bibinfo{year}{2020}).

\bibitem{dommermuth1987high}
\bibinfo{author}{Dommermuth, D.~G.} \& \bibinfo{author}{Yue, D.~K.}
\newblock \bibinfo{title}{A high-order spectral method for the study of
  nonlinear gravity waves}.
\newblock \emph{\bibinfo{journal}{Journal of Fluid Mechanics}}
  \textbf{\bibinfo{volume}{184}}, \bibinfo{pages}{267--288}
  (\bibinfo{year}{1987}).

\bibitem{west1987new}
\bibinfo{author}{West, B.~J.}, \bibinfo{author}{Brueckner, K.~A.},
  \bibinfo{author}{Janda, R.~S.}, \bibinfo{author}{Milder, D.~M.} \&
  \bibinfo{author}{Milton, R.~L.}
\newblock \bibinfo{title}{A new numerical method for surface hydrodynamics}.
\newblock \emph{\bibinfo{journal}{Journal of Geophysical Research: Oceans}}
  \textbf{\bibinfo{volume}{92}}, \bibinfo{pages}{11803--11824}
  (\bibinfo{year}{1987}).

\bibitem{HOS-NWT}
\bibinfo{author}{{Ecole Centrale Nantes, LHEEA}}.
\newblock \bibinfo{title}{Open-source release of {HOS-NWT}}.
\newblock \bibinfo{note}{\url{https://github.com/LHEEA/HOS-NWT}}.

\bibitem{bonnefoy2010time}
\bibinfo{author}{Bonnefoy, F.}, \bibinfo{author}{Ducrozet, G.},
  \bibinfo{author}{Le~Touz{\'e}, D.} \& \bibinfo{author}{Ferrant, P.}
\newblock \bibinfo{title}{Time domain simulation of nonlinear water waves using
  spectral methods}.
\newblock In \emph{\bibinfo{booktitle}{Advances in numerical simulation of
  nonlinear water waves}}, \bibinfo{pages}{129--164} (\bibinfo{publisher}{World
  Scientific}, \bibinfo{year}{2010}).

\bibitem{ducrozet2012modified}
\bibinfo{author}{Ducrozet, G.}, \bibinfo{author}{Bonnefoy, F.},
  \bibinfo{author}{Le~Touz{\'e}, D.} \& \bibinfo{author}{Ferrant, P.}
\newblock \bibinfo{title}{A modified high-order spectral method for wavemaker
  modeling in a numerical wave tank}.
\newblock \emph{\bibinfo{journal}{European Journal of Mechanics-B/Fluids}}
  \textbf{\bibinfo{volume}{34}}, \bibinfo{pages}{19--34}
  (\bibinfo{year}{2012}).

\end{thebibliography}
\end{document}